\documentclass[12pt,a4paper]{article}

\usepackage{epsfig}
\usepackage{amsmath,amsfonts,amssymb}
\usepackage{cite}
\usepackage{colortbl}
\usepackage{pifont}

\providecommand{\openone}{\leavevmode\hbox{\small1\kern-3.8pt\normalsize1}}

\newcommand{\swz}{\sigma_{WZ}}
\newcommand{\szz}{\sigma_{ZZ}}
\newcommand{\sww}{\sigma_{WW}}
\newcommand{\shh}{\sigma_{HH}}
\newcommand{\swh}{\sigma_{WH}}
\newcommand{\szh}{\sigma_{ZH}}
\newcommand{\et}{\mbox{${E_T}$}}
\newcommand{\etmiss}{\mbox{$\protect \raisebox{.3ex}{$\not$}\et$}}

\begin{document}

\begin{center}
\begin{Large}
{\bf Triboson interpretations of the ATLAS \\[1mm] diboson excess}
\end{Large}

\vspace{0.5cm}
J.~A.~Aguilar--Saavedra \\[1mm]
\begin{small}
{Departamento de F\'{\i}sica Te\'orica y del Cosmos, 
Universidad de Granada, 18071 Granada, Spain} 
\end{small}
\end{center}

\begin{abstract}
The ATLAS excess in fat jet pair production is kinematically compatible with the decay of a heavy resonance into two gauge bosons plus an extra particle. This hypothesis would explain the absence of such a large excess in the analogous CMS analysis of fat dijet final states, as well as the negative results of diboson resonance searches in the semi-leptonic decay modes. If the extra particle is the Higgs boson, this hypothesis might also explain---statistical fluctuations aside---why the CMS search for $WH$ resonances in the semi-leptonic channel finds some excess while in the fully hadronic one it does not have a significant deviation.
\end{abstract}
\section{Introduction}

The ATLAS Collaboration has recently observed a localised excess in the invariant mass distribution of pairs of fat jets, hereafter denoted by $J$, around $m_{JJ} \simeq 2$ TeV~\cite{Aad:2015owa}. Fat jets can be produced in the hadronic decay of boosted bosons $V=W,\,Z$, where the two quarks from the boson decay merge into a single jet. Using jet substructure analyses, the fat jets are tagged as resulting from a boson decay. In addition, in ref.~\cite{Aad:2015owa} the jets are identified as $W$ or $Z$ bosons if the jet mass $m_J$ satisfies $|m_J - M_W| \leq 13$ GeV or $|m_J - M_Z| \leq 13$ GeV, respectively.
The excess in the $m_{JJ}$ spectrum appears for $WZ$, $ZZ$ and $WW$ selections, with statistical significances of $3.4\,\sigma$, $2.9\,\sigma$ and $2.6\,\sigma$.\footnote{Notice that $J$ can be simultaneously tagged as $W$ and $Z$ with these criteria, as the mass windows for $W$ and $Z$ tagging partially overlap. This indicates, in particular, that a $WZ$ signal can yield significant excesses in the $WW$ and $ZZ$ selections too.}
These three channels are not independent and some events fall into two or even the three above categories. A statistical combination of the three channels must take this fact into account, and has not been yet performed.

$W$ and $Z$ bosons are known to decay leptonically, therefore a potential diboson resonance should also show up in (semi-)leptonic channels. But it does not. For example, for spin-1 $WZ$ resonances with a mass $M= 2$ TeV the ATLAS Collaboration has the following 95\% confidence level (CL) upper limits on the cross section $\swz$:
\begin{itemize}
\item[(i)] $WZ \to J J$ channel~\cite{Aad:2015owa}: $\swz < 30$ fb, with an expected limit of 12 fb. This corresponds to a  $3.4\,\sigma$ excess. As mentioned above, this search is also sensitive to $WW$ and $ZZ$ resonances.
\item[(ii)] $WZ \to \ell \nu J$ channel, $\ell=e,\mu$~\cite{Aad:2015ufa}: $\sigma_{WZ} < 9.5$ fb (11 fb expected). This search is also sensitive to $WW$ resonances.
\item[(iii)] $WZ \to \ell \ell J $ channel~\cite{Aad:2014xka}: $\swz < 20$ fb (16 fb expected). This search is also sensitive to $ZZ$ resonances.
\item[(iv)] $WZ \to 3\ell \nu$ channel~\cite{Aad:2014pha}: $\swz < 22$ fb (24 fb expected).
\end{itemize}
Therefore, the absence of any signal in the semi-leptonic channels excludes a significant excess in the fully hadronic one. In particular, the $\ell \nu J$ channel has better sensitivity than the $JJ$ one---the expected limit is smaller---but deviations from the Standard Model (SM) predictions are not found. (We note in passing that the ``definition'' of the fat jet $J$ is the same in the three ATLAS analyses~\cite{Aad:2015owa,Aad:2015ufa,Aad:2014xka}.) The tension between the ATLAS $JJ$ and $\ell \nu J$ searches can be quantified with a simple event counting. With the $WZ$ selection, there are 15 observed events in the hadronic channel with $m_{JJ} \in [1.85,2.15]$ TeV, for an expected background of approximately 7 events. With the colected luminosity of 20.3 fb$^{-1}$, these eight extra events require a signal cross section $\sigma_{WZ}^\text{[peak]}$ times branching ratio and efficiency factors of
\begin{equation}
\sigma_{WZ}^\text{[peak]} \times \text{Br}(W \to q \bar q') \times \text{Br}(Z \to q \bar q) \times \text{eff} = 0.39 ~\text{fb}\,.
\end{equation}
The superscript in $\sigma_{WZ}^\text{[peak]}$ emphasises that this is the cross section for $m_{JJ} \in [1.85,2.15]$ TeV alone. (Because the reconstructed invariant mass distribution of a 2 TeV resonance is wider than the invariant mass interval considered, the actual $WZ$ cross section $\sigma_{WZ}$ required to reproduce the excess is around twice larger.) For the selection efficiency of 0.14 given in ref.~\cite{Aad:2015owa}, which does not include the hadronic branching ratios, we find $\sigma_{WZ}^\text{[peak]} = 6$ fb. Now let us reverse the procedure to estimate the number of extra events that should show up in the $\ell \nu J$ channel. With the efficiency of 0.25 for $WZ \to \ell \nu q \bar q$ given in ref.~\cite{Aad:2015ufa}, where $\ell$ includes electrons, muons and taus, $\sigma_{WZ}^\text{[peak]} = 6$ fb would yield 7 extra events in the mass interval $m_{\ell \nu J} \in [1.8,2.1]$ TeV, which practically coincides with the interval considered for the hadronic channel. But there are six observed events in this mass interval, for an expected background around 6.5 events. This would amount to a $2.4\,\sigma$ underfluctuation of the signal observed in the $JJ$ channel.

The CMS Collaboration has also looked for spin-1 $WZ$ resonances in the $JJ$ channel~\cite{Khachatryan:2014hpa}, using a slightly different strategy. For a resonance mass $M= 2$ TeV, the limit is $\swz < 12$ fb, with an expected limit of 8 fb. Although there is some excess around $M=2$ TeV, the CMS upper limit on the cross section is also in tension with the interpretation of the ATLAS excess as a narrow $WZ$ resonance. The CMS analysis of the semi-leptonic channel~\cite{Khachatryan:2014gha} does not observe any excess at $M=2$ TeV.\footnote{In the $\ell \ell J$ mode a $2\,\sigma$ deviation is found at invariant masses around 1.8 TeV, though limits are not reported for spin-1 particles. For a 1.8 TeV spin-2 graviton the limits are $\szz < 15$ fb, with 7 fb expected, but these numbers cannot be directly compared to the ATLAS limits~\cite{Aad:2014xka} because of the different signal efficiencies.}
The CMS search in the fully leptonic channel~\cite{Khachatryan:2014xja} gives an upper limit $\swz <  20$ fb for $M=2$ TeV.

More generally, a mixture of spin-one $WZ$, $ZZ$ and $WW$ resonances with nearly the same mass (they can appear for example in models with a heavy $\text{SU}(2)_L$ vector boson triplet~\cite{deBlas:2012qp}) cannot explain the $JJ$ excess. This can be shown with a simple exercise. Since the jet mass cuts in the ATLAS semi-leptonic searches, $65 < m_J < 105~\text{GeV}$~\cite{Aad:2015ufa} and $70 < m_J < 110~\text{GeV}$~\cite{Aad:2014xka}, are wide enough to accept $W$ and $Z$ bosons, one can approximately rewrite the corresponding 95\% CL cross section limits in (ii) and (iii) as
\begin{align}
& \swz + 0.96 \,\sww < 9.5~\text{fb} \notag \, \\
& \swz + 1.04 \,\szz < 20~\text{fb} \,,
\label{ec:reint}
\end{align}
by rescaling with the $W$ and $Z$ hadronic branching ratios and ignoring small efficiency differences that may arise from the different $W$ and $Z$ boson masses. For the fully hadronic final state, a relative acceptance factor $\sim 0.8$ for $WW$ and $ZZ$ diboson signals with the $WZ$ selection can be roughly estimated from the overlap of the jet mass distribution for signal samples in Ref.~\cite{Aad:2015owa}. The total $JJ$ signal $\sigma_\text{tot}$ with the $WZ$ selection can then be estimated as
\begin{equation}
\sigma_\text{tot} = \swz + 0.8 \, \szz + 0.8 \, \sww \,,
\end{equation}
and reaches a maximum $\sigma_\text{tot} \lesssim 23$ fb given the constraints in Eqs.~(\ref{ec:reint}). Besides, the CMS search in the $JJ$ channel with a wide mass window $70 < m_{JJ} < 100$ GeV does not observe such a large excess, as mentioned above.
One can also wonder if a different type of resonance ({\it e.g.} with spin 2) yielding different diboson helicity configurations may have an efficiency in the semi-leptonic channels much smaller than for a vector resonance ---which mainly produces longitudinal vector bosons--- therefore turning the non-observation of a large signal compatible with the $JJ$ excess. As it will be shown, this is not the case because the event selection criteria used by the ATLAS and CMS Collaborations mainly focus on the kinematics of the reconstructed bosons rather than on its leptonic decay products. 

The possibility that the ATLAS $JJ$ anomaly is due to Higgs production incorrectly tagged as $W/Z$ is strongly disfavoured by other measurements:
\begin{itemize}
\item[(v)] An ATLAS search for $HH$ resonances~\cite{Aad:2015uka} gives $\shh < 30$ fb for $M=2$ TeV, with an expected limit around 45 fb.
\item[(vi)] A search for $ZH$ and $WH$ resonances in the fully hadronic $JJ$ channel by the CMS Collaboration~\cite{Khachatryan:2015bma} yields cross section limits $\szh < 7$ fb and $\swh < 7$ fb, very close to the expected ones. A preliminary $WH$ resonance search in the $\ell \nu J$ final state~\cite{CMS:2015gla} yields a $2.2 \, \sigma$ excess at $m_{WH} = 1.8$ TeV, with  $\swh \lesssim 40$ fb for an expected limit of 20 fb. But the hypothesis of a $WH$ resonance behind the latter excess is disfavoured by the former  search, which gives $\swh < 20$ fb for $m_{WH}=1.8$ TeV. A similar analysis by the ATLAS Collaboration~\cite{Aad:2015yza} is less sensitive, giving $\szh \lesssim 15$ fb, $\swh \lesssim 35$ fb, for a resonance mass $M= 1.9$ TeV.
\end{itemize}
Since neither $HH$ nor $VH$ resonance signals show up in these dedicated searches, it is very unlikely that they could contribute significantly to the $JJ$ excess with a fat dijet selection optimised for $VV$ production. Additionally, one expects relations between $VV$ and $VH$ decay fractions of heavy resonances in definite models~\cite{Hisano:2015gna}. All this overwhelming set of related SM-like measurements has motivated the caution by the ATLAS Collaboration regarding this excess, but it has not discouraged early interpretations as new diboson resonances of technicolour models~\cite{Fukano:2015hga,Franzosi:2015zra}. While ref.~\cite{Fukano:2015hga} only takes into account the limit on the production of $WZ$ resonances from the fully leptonic channel (the weakest one), ref.~\cite{Franzosi:2015zra} attributes the tension among the searches in different $W,Z$ decay channels to statistics. Other $W'/Z'$ interpretations~\cite{Cheung:2015nha} only focus on the $JJ$ excess overlooking the null results obtained in the other decay modes of the gauge boson pair.

Statistical fluctuations aside, experimental data seem to disfavour the possibility that the ATLAS $JJ$ excess results from a diboson resonance. We are then led to consider that, it this excess is real, it might be due to something different that looks as a diboson peak due to the kinematical selection applied to reduce SM backgrounds. As we will show in this paper, a requirement on transverse momenta applied in ref.~\cite{Aad:2015owa} shapes certain resonant $VVX$ signals, with $X$ an extra particle, making them look like a $VV$ resonance. Such a requirement is not used by the corresponding analysis of the $JJ$ final state by the CMS Collaboration~\cite{Khachatryan:2014hpa}, nor in the ATLAS analyses of semi-leptonic final states. In section~\ref{sec:2} we explore several final state topologies to find in which cases  a diboson peak is kinematically produced---without an actual diboson resonance. In section~\ref{sec:3} we show in two benchmark examples how a possible signal would look like with the ATLAS and CMS fat dijet selections, as well as in the ATLAS analyses of semi-leptonic final states. We present our conclusions in section~\ref{sec:4}.
In an appendix we discuss to what extent the event selection efficiencies in the semi-leptonic channels depend on the different diboson helicities.

\section{Alternative topologies for the excess}
\label{sec:2}

In this section we explore different topologies in which a diboson pair plus an extra particle $X$ are produced, focusing for definiteness on $WZ$ production. (Results are the same for $WW$ and $ZZ$, obviously.) In principle, the $X$ particle could either be invisible (thus a potential dark matter candidate), or a new relatively light scalar with dominant hadronic decay, or even just a SM gauge or Higgs boson. We work at the partonic level, calculating matrix elements for processes with new generic scalars, fermions or vector bosons, integrating over phase space and parton distribution functions, and examining how the $VV$ invariant mass distribution is shaped by the following cuts applied in the ATLAS dijet analysis~\cite{Aad:2015owa}:
\begin{enumerate}
\item Boson pseudo-rapidity $|\eta_{1,2}| \leq 2$ and rapidity difference $|y_1-y_2| < 1.2$, where the indices 1 and 2 denote the two bosons.
\item At least one of the two bosons must have transverse momentum $p_T$ greater than 360 GeV.
\item Transverse momentum asymmetry $|p_T^1-p_T^2| / (p_T^1+p_T^2) < 0.15$. Together with the rapidity difference cut, this requirement selects approximately back-to-back bosons even if they are not produced from the decay of an $s$-channel resonance.
 \end{enumerate}
In this section we do not impose the requirement on missing energy $\etmiss < 350$ GeV of the analysis in ref.~\cite{Aad:2015owa} to remain as general as possible, since in principle $X$ could be invisible or decay hadronically. At any rate, the application of such a cut does not significantly modify the obtained distributions.

Matrix elements are evaluated using {\sc HELAS}~\cite{Murayama:1992gi} including the decay of the $W$ and $Z$ bosons. Phase space integration is done by implementing these processes into the generator {\sc Protos}~\cite{protos}. We give our results for specific choices of the new particles, {\it e.g.} assuming that $X$ is a neutral scalar. But our results are more general, as they are mainly based on the kinematics of cascade decays, and we have explicitly checked this fact by using ``flat'' matrix elements, constant except for the resonant propagators, which make no assumption on the spin or charge of the new particles. Results are also independent of the incoming partons. For definiteness we assume $u \bar d$, $\bar u d$ initial states for the production of $W^+ Z X$, $W^- Z X$, respectively. However, the presence of the extra particle $X$ opens the possibility of gluon-initiated processes with larger partonic luminosities.

\subsection{Non-resonant $VV$ and $VVX$ production}
\label{sec:2.1}

\begin{figure}[h!]
\begin{center}
\begin{tabular}{ccc}
\includegraphics[height=2.5cm,clip=]{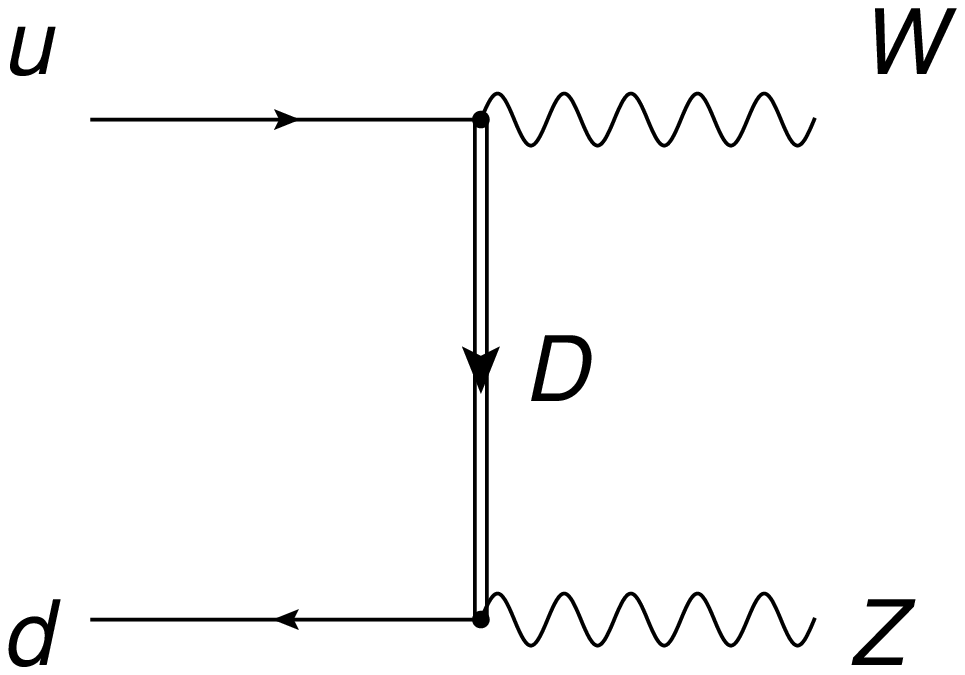} & \quad &
\includegraphics[height=2.5cm,clip=]{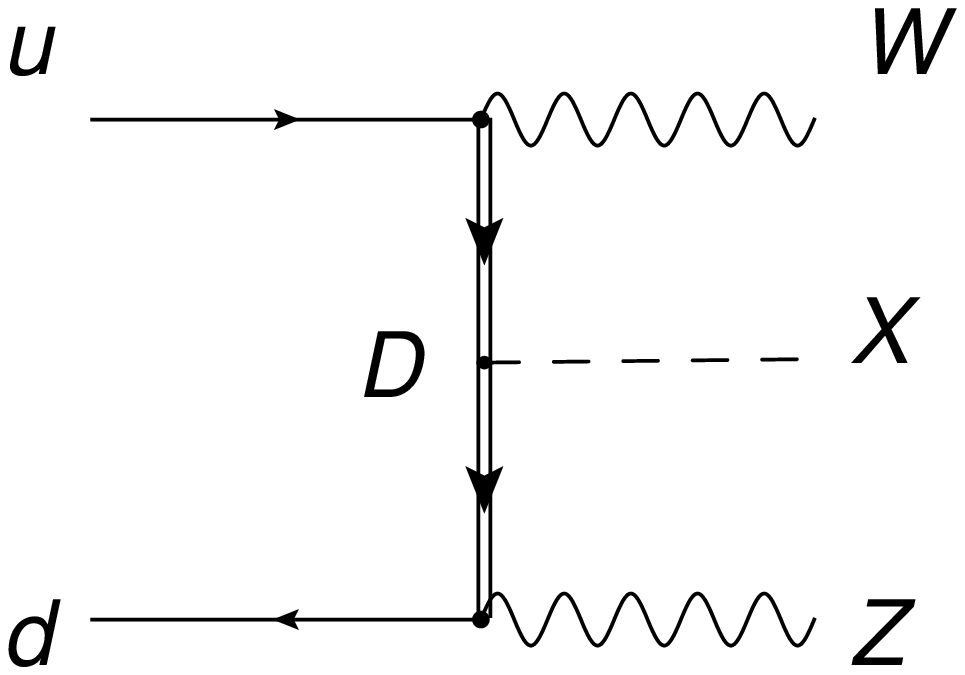}
\end{tabular}
\caption{Sample diagrams for $WZ$ and $WZX$ production, with $X$ a neutral scalar.}
\label{fig:topVV}
\end{center}
\end{figure}

\begin{figure}[htb]
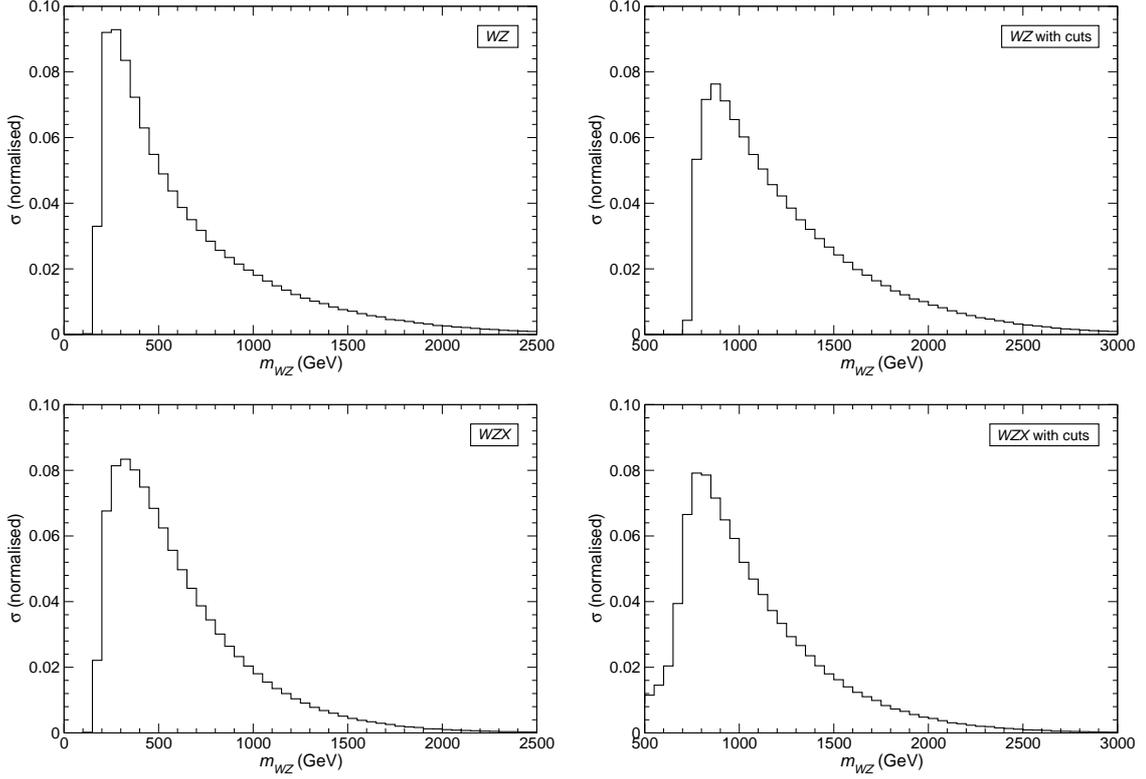

\begin{center}
\begin{tabular}{cc}
\includegraphics[height=5.cm,clip=]{Figs/dist-VV} &
\includegraphics[height=5.cm,clip=]{Figs/dist-VVc} \\[1mm]
\includegraphics[height=5.cm,clip=]{Figs/dist-VVX} &
\includegraphics[height=5.cm,clip=]{Figs/dist-VVXc} 
\end{tabular}
\caption{Diboson invariant mass distribution for non-resonant $WZ$ (up) and $WZX$ (down) production without cuts (left) and after cuts (right).}
\label{fig:distVV}
\end{center}
\end{figure}

It is clear from the beginning that non-resonant diboson production, with or without an extra particle, cannot give a peak at an invariant mass as high as 2 TeV merely with the application of the kinematical cuts in (1--3). However, it is interesting to consider this academic case to investigate how an unaccounted SM contribution could be affected by this event selection. We consider $u \bar d \to W^+ Z$, $u \bar d \to W^+ Z X$ (with $X$ a neutral scalar) plus the charge conjugate processes, mediated by a $t$-channel heavy quark $D$, as shown in figure~\ref{fig:topVV}. The normalised $WZ$ invariant mass distributions before and after cuts are presented in figure~\ref{fig:distVV} (top). For $WZX$ we take a scalar mass $M_X=100$ GeV, but the results are rather independent of this value. The cuts reduce the cross section by a factor of 5 for $WZ$ and 11 for $WZX$ and in both cases they maintain the shape of the distribution, with a shift towards larger invariant masses and a long tail.

\subsection{Resonant $VVX$ production}
\label{sec:2.2}

\begin{figure}[htb]
\begin{center}
\begin{tabular}{ccccc}
\includegraphics[height=2.2cm,clip=]{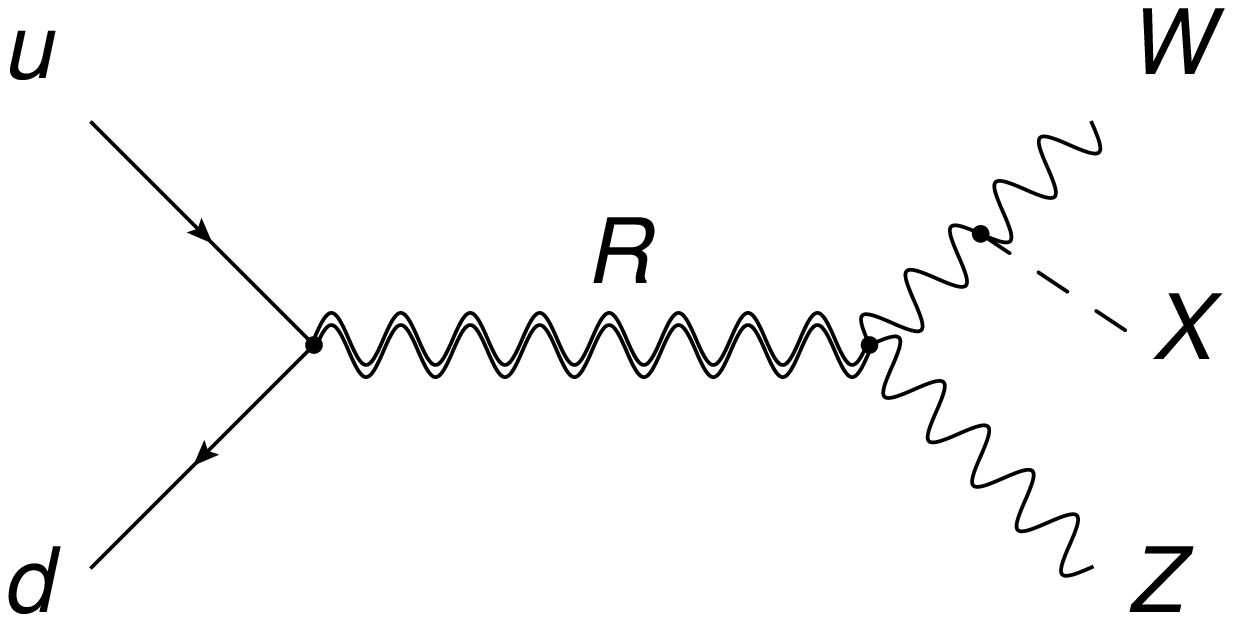} & \quad &
\includegraphics[height=2.2cm,clip=]{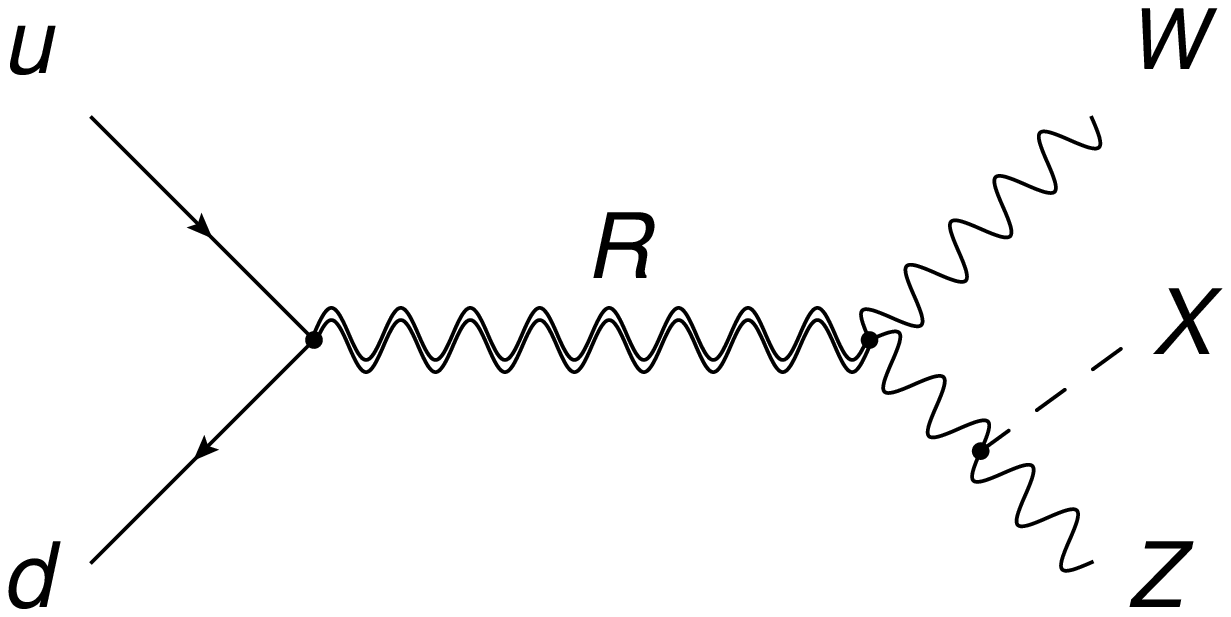} & \quad &
\includegraphics[height=2.2cm,clip=]{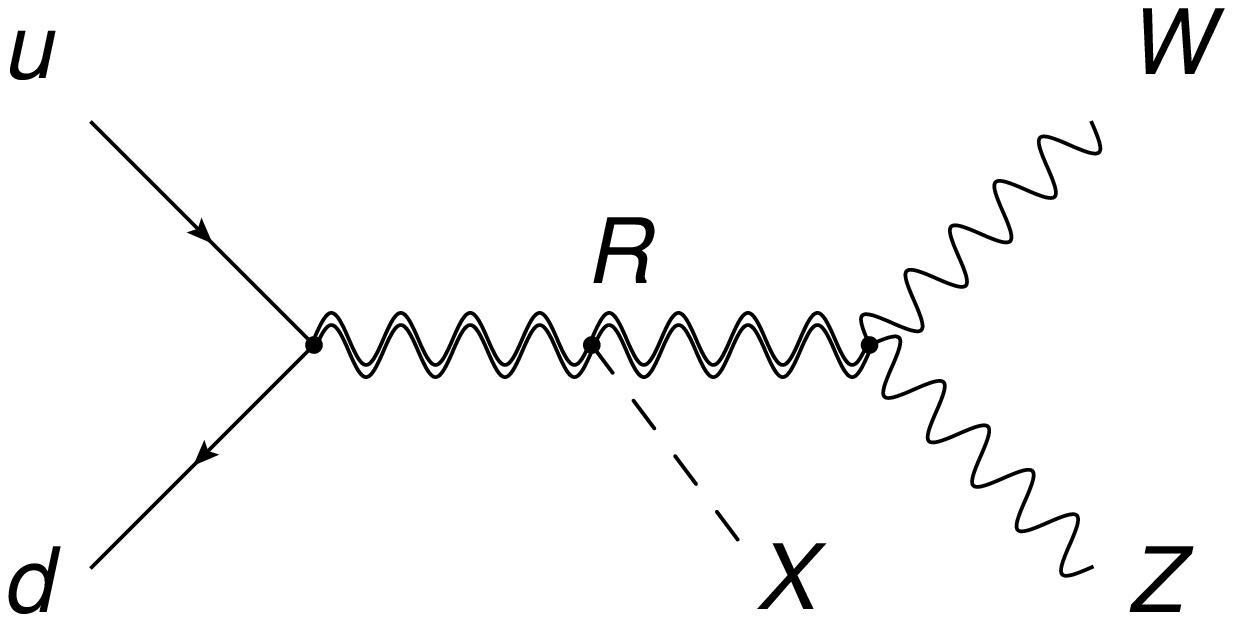} \\
(a) & & (b) & & (c) \\
\includegraphics[height=2.2cm,clip=]{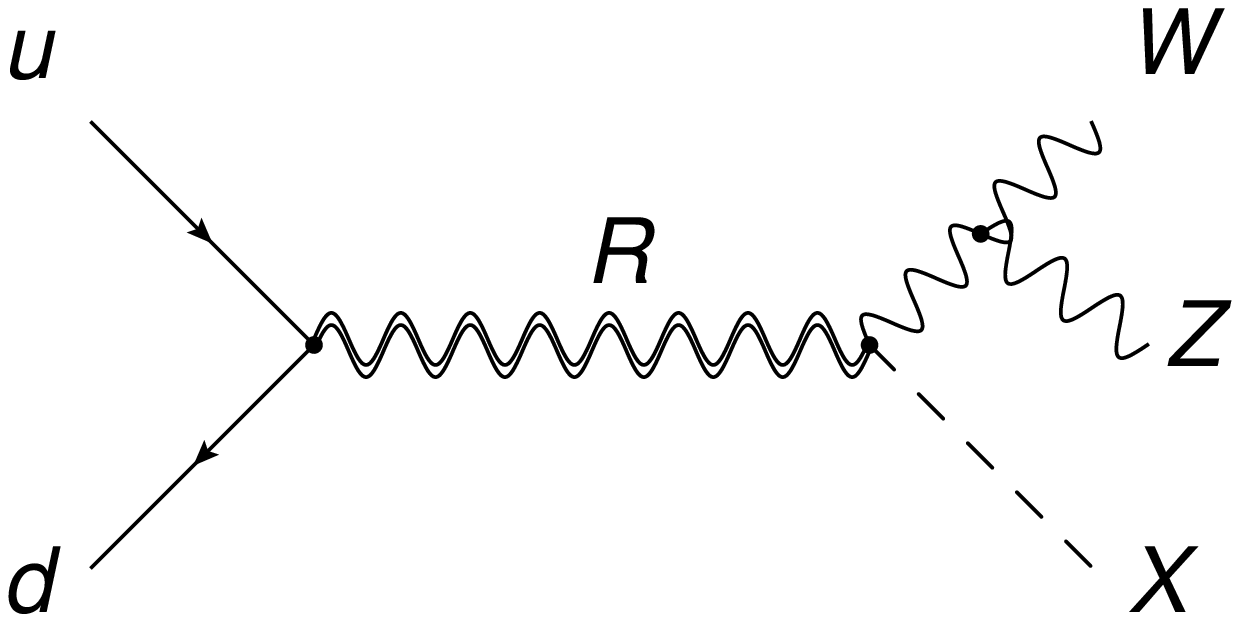} & \quad &
\includegraphics[height=2.2cm,clip=]{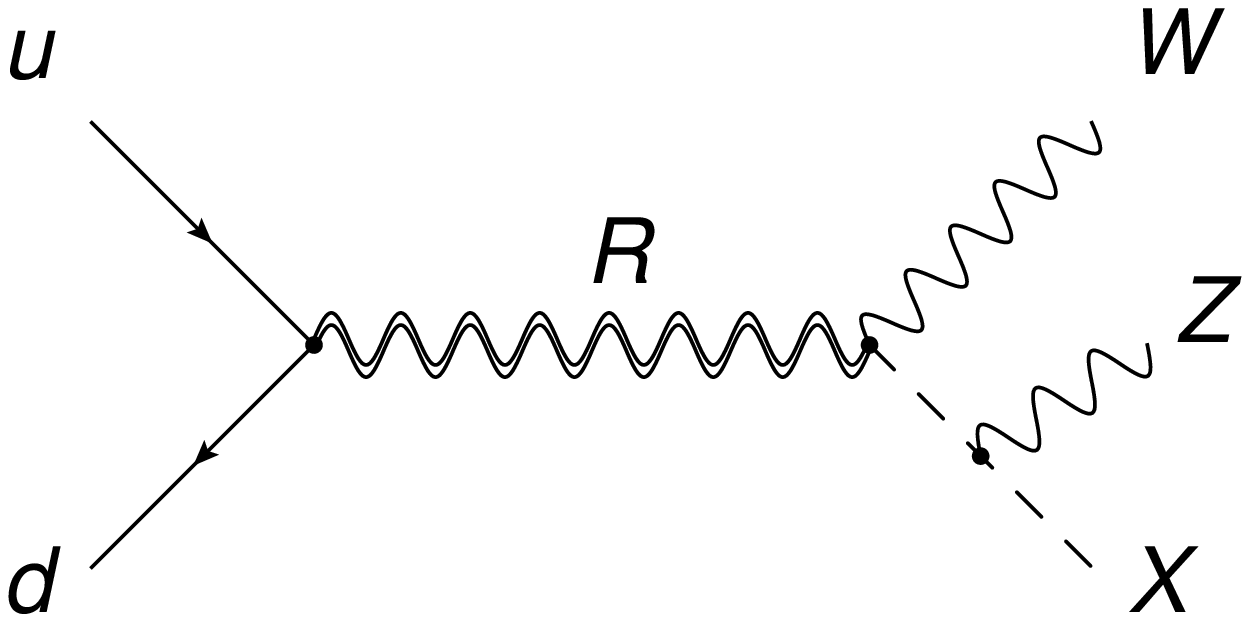} & \quad &
\includegraphics[height=2.2cm,clip=]{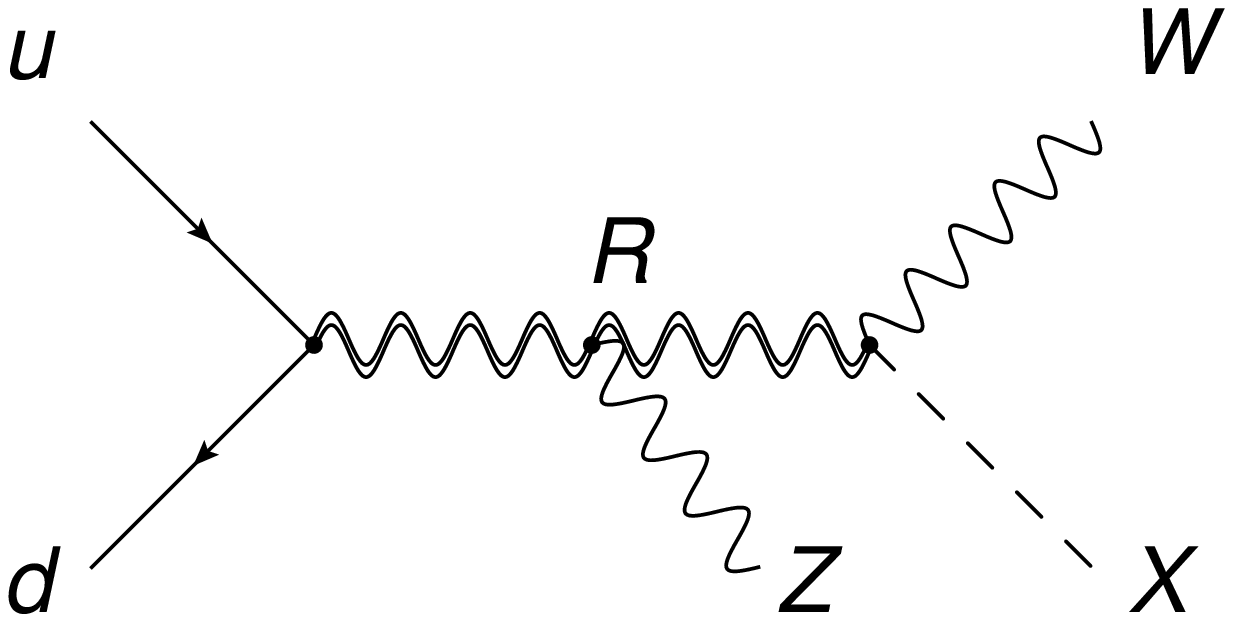} \\
(d) & & (e) & & (f)
\end{tabular}
\caption{Sample diagrams for $R \to WZX$ production (without an extra intermediate state), assuming $X$ is a neutral scalar.}
\label{fig:topVVX}
\end{center}
\end{figure}

A heavy resonance $R$ can decay into $WZX$ as shown in figure~\ref{fig:topVVX}, where we assume for definiteness that $X$ is a neutral scalar and $R$ a charged vector boson. Diagrams (a--c) require $R \to WZ$ without the extra particle $X$, {\it i.e.} $R$ is a diboson resonance, which we do not consider in this work as argued in the introduction. Diagrams (d--f) are sub-leading with respect to  $R \to WX$ and these processes are expected to be small. Otherwise, these topologies give results similar to the ones with an extra intermediate particle $Y$, studied in the next subsection. We omit a detailed study for brevity.

\subsection{Resonant $VY \to VVX$ production}
\label{sec:2.3}

\begin{figure}[t]
\begin{center}
\begin{tabular}{ccc}
\includegraphics[height=2.2cm,clip=]{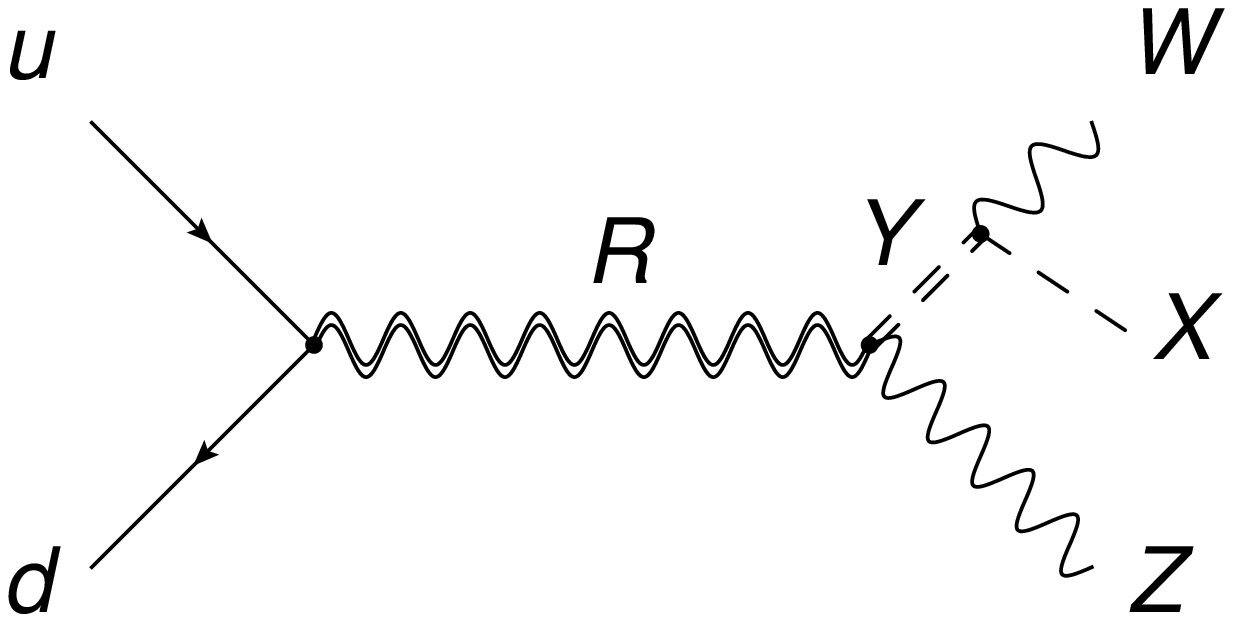} & \quad &
\includegraphics[height=2.2cm,clip=]{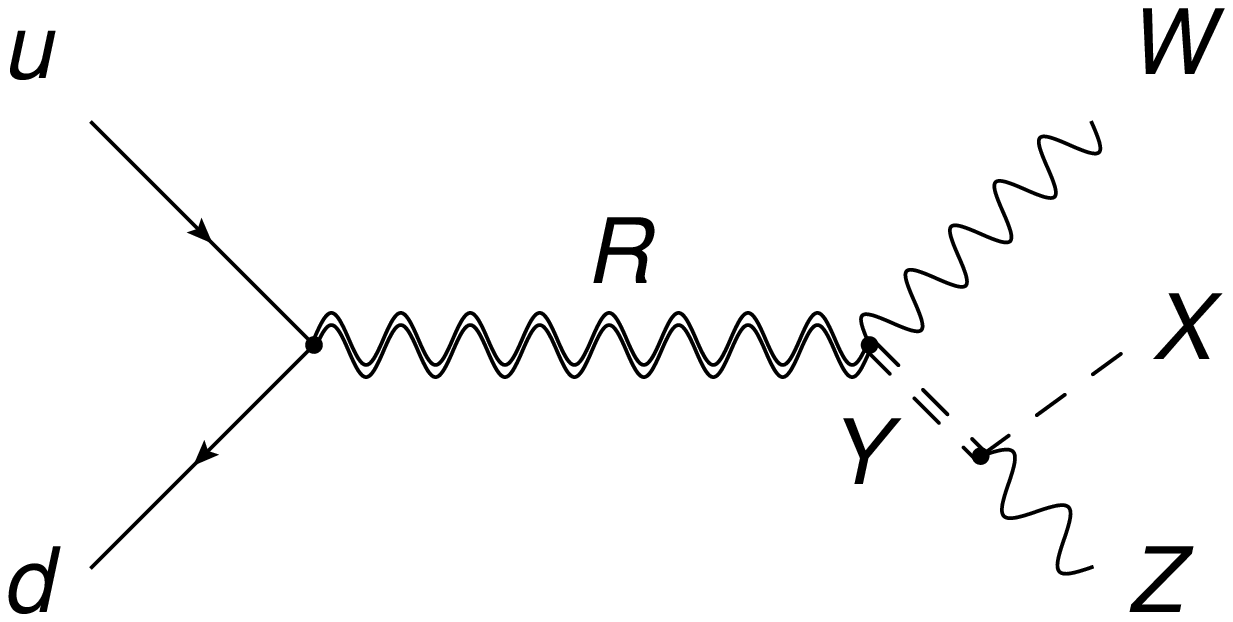}  \\
(a) & & (b) 
\end{tabular}
\caption{Sample diagrams for $R \to VY \to VVX$ production, with $X$ a neutral scalar.}
\label{fig:topVY}
\end{center}
\end{figure}

\begin{figure}[htb]
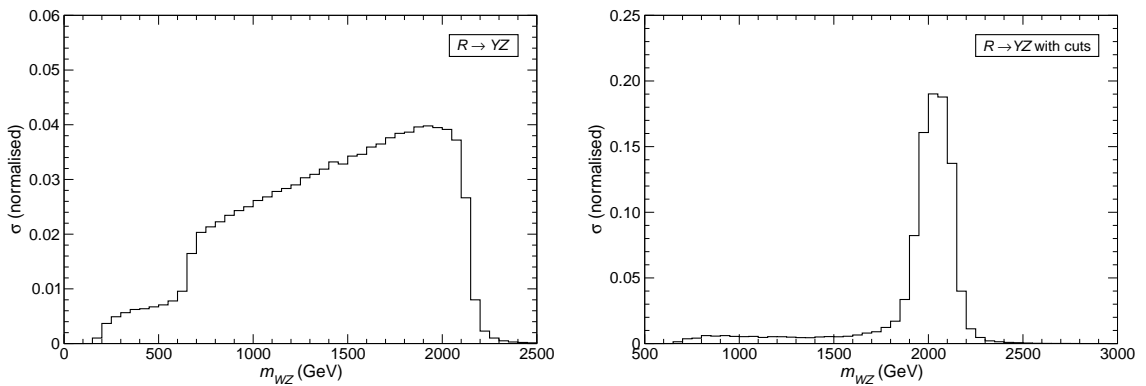

\begin{center}
\begin{tabular}{cc}
\includegraphics[height=5.cm,clip=]{Figs/dist-VY} &
\includegraphics[height=5.cm,clip=]{Figs/dist-VYc}
\end{tabular}
\caption{Diboson invariant mass distribution in $R \to YZ \to WZX$ production without cuts (left) and after cuts (right).}
\label{fig:distVY}
\end{center}
\end{figure}

A heavy resonance $R$ can also decay into $WZX$ via an intermediate on-shell state $Y$, as represented in figure~\ref{fig:topVY}.
(Let us also mention for completeness that non-resonant production $VY \to VVX$ does not produce a peak but gives distributions similar to those in figure~\ref{fig:distVV}.) For definiteness, we have taken $Y$ to be a scalar but our conclusions are independent of this choice, and independent of its mass $M_Y$ to a large extent. As an example we take the decay chain in figure~\ref{fig:topVY} (a) with $M_R = 2.3$ TeV, $\Gamma_R = 50$ GeV (chosen to reproduce a peak around 2 TeV), $M_Y = 300$ GeV, $\Gamma_Y = 5$ GeV, and $M_X = 100$ GeV. Upon application of the kinematical cuts, which reduce the cross section by a factor of 7, the wide distribution in figure~\ref{fig:distVY} (left) adopts a very peaked shape, see the right panel. The cascade decay $R \to VY \to VVX$ is then a suitable candidate to explain why the ATLAS Collaboration observes a peak structure in fat dijet searches while the CMS Collaboration has a smaller excess. The decay chain in diagram (b) gives identical results.

 This signal shaping can be understood by writing the squared diboson invariant mass as
\begin{equation}
m_{WZ}^2 = M_X^2 + M_R \left( M_R - 2 \sqrt{M_X^2 + q^2} \right) \,,
\end{equation}
with $q$ the modulus of the three-momentum of the $X$ particle in the centre-of-mass (CM) frame. The maximum value $m_{WZ} = M_R - M_X$ is reached for $q=0$. The requirement of central dibosons with similar transverse momentum selects the kinematical configurations with $q \sim 0$, therefore making $m_{WZ}$ close to $M_R$. This happens quite independently of $M_Y$, and we have checked that for $M_Y = 1$ TeV a peak structure still appears after application of the kinematical cuts.

\subsection{Resonant $YY \to VVXX$ production}
\label{sec:2.4}

\begin{figure}[htb]
\begin{center}
\includegraphics[height=2.2cm,clip=]{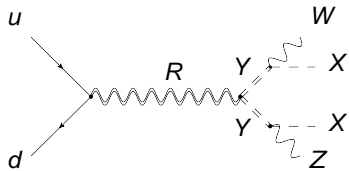}
\caption{Sample diagram for $R \to YY \to VVXX$ production, with $X$ a neutral scalar.}
\label{fig:topYY}
\end{center}
\end{figure}

\begin{figure}[htb]
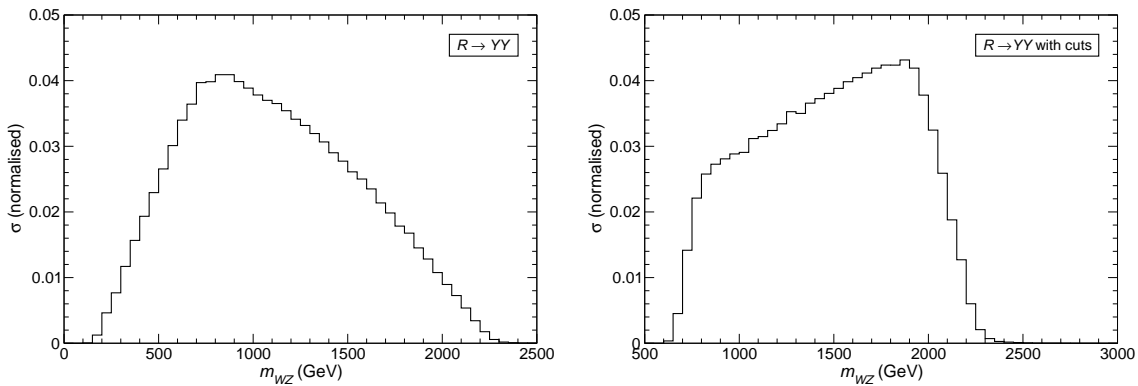

\begin{center}
\begin{tabular}{cc}
\includegraphics[height=5.cm,clip=]{Figs/dist-YY} &
\includegraphics[height=5.cm,clip=]{Figs/dist-YYc}
\end{tabular}
\caption{Diboson invariant mass distribution in $R \to YZ \to WZX$ production without cuts (left) and after cuts (right).}
\label{fig:distYY}
\end{center}
\end{figure}

For the sake of completeness we have also investigated a process where two extra particles are produced, $R \to YY \to VVXX$, as depicted in figure~\ref{fig:topYY}. The intermediate resonances $Y$ are neutral or charged, and their masses are assumed equal. The requirement of back-to-back dibosons does not fix their invariant mass in this case because there are more degrees of freedom. We take $M_R = 2.6$ TeV, $\Gamma_R = 50$ GeV, $M_Y = 300$ GeV, $\Gamma_Y = 5$ GeV, $M_X = 100$ GeV. Figure~\ref{fig:distYY} shows the diboson invariant mass distribution before and after the kinematical cuts, which reduce the cross section by a factor of 5. This topology does not seem so promising because the shape of the signal after kinematical cuts is not really a peak.

\section{Triboson interpretations}
\label{sec:3}

We further investigate the $R \to YZ \to WZX$ topology with a fast detector simulation using {\sc PGS4}~\cite{pgs4}. This exercise is not intended to provide a detailed description of the possible signals for comparison with data, which requires an implementation of jet filtering~\cite{Butterworth:2008iy} and pruning~\cite{Ellis:2009su}, used by the ATLAS and CMS Collaborations, respectively. Such a detailed study is beyond the scope of this work.
Rather, our aim is to test whether a localised excess in the ATLAS dijet search is indeed compatible with the absence of such peaks in other analyses. Among those, we consider:
\begin{itemize}
\item The ATLAS $\l\nu J$ analysis,  with the event selection criteria of (i) exactly one charged lepton $\ell$: either an electron with  $p_T > 20$ GeV and rapidity $|\eta| < 1.37$ or $1.52 < |\eta| < 2.47$; or a muon with $p_T > 20$ GeV and $|\eta| < 2.5$; (ii) a fat jet with $p_T > 400$ GeV, $|\eta| < 2$; (iii) $\etmiss > 30$ GeV, with a difference of azimuthal angle $\Delta \phi(\etmiss,J) > 1$ with the jet; (iv) $p_T^W > 400$ GeV, where the $W$ boson momentum is reconstructed with the charged lepton and neutrino ($\etmiss$) momentum, imposing the on-shell condition. An additional requirement of no $b$-tagged jets is not considered since it only affects the signal if $X$ decays into $b$ quarks.
\item The ATLAS $\ell \ell J$ analysis, requiring (i) exactly two same-flavour charged leptons within the above acceptance, with invariant mass $66 < m_{\ell \ell} < 116$ GeV; (ii) a fat jet with $p_T > 400$ GeV, $|\eta| < 1.2$; $p_T^Z > 400$ GeV, where the $Z$ boson momentum is reconstructed from the two charged lepton momenta.
\item The CMS $JJ$ analysis, which selects two jets with $p_T > 30$ GeV and $|\eta| < 2.5$, separation $|\Delta \eta| < 1.3$ and invariant mass $m_{JJ} > 890$ GeV.
\end{itemize}
For the ATLAS dijet analysis we apply the kinematical cuts (1--3) listed in section~\ref{sec:2} to the two leading jets, plus the requirements of no charged leptons in the above ATLAS common acceptance region, and $\etmiss < 350$ GeV.

\begin{figure}[htb]
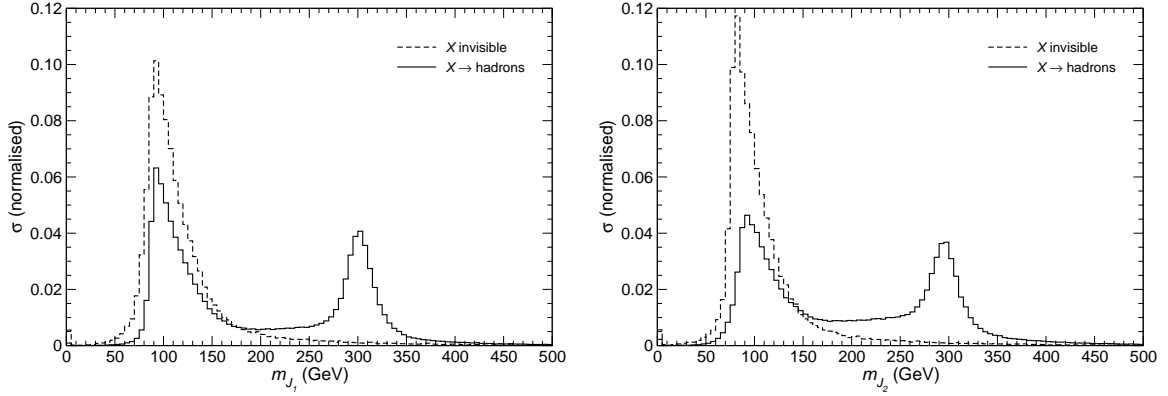

\begin{center}
\begin{tabular}{cc}
\includegraphics[height=5.2cm,clip=]{Figs/mJ1} &
\includegraphics[height=5.2cm,clip=]{Figs/mJ2} 
\end{tabular}
\caption{Mass of the leading (left) and sub-leading (right) jets in the two benchmarks, obtained in the simulation. }
\label{fig:jetmass}
\end{center}
\end{figure}

\begin{figure}[htb]
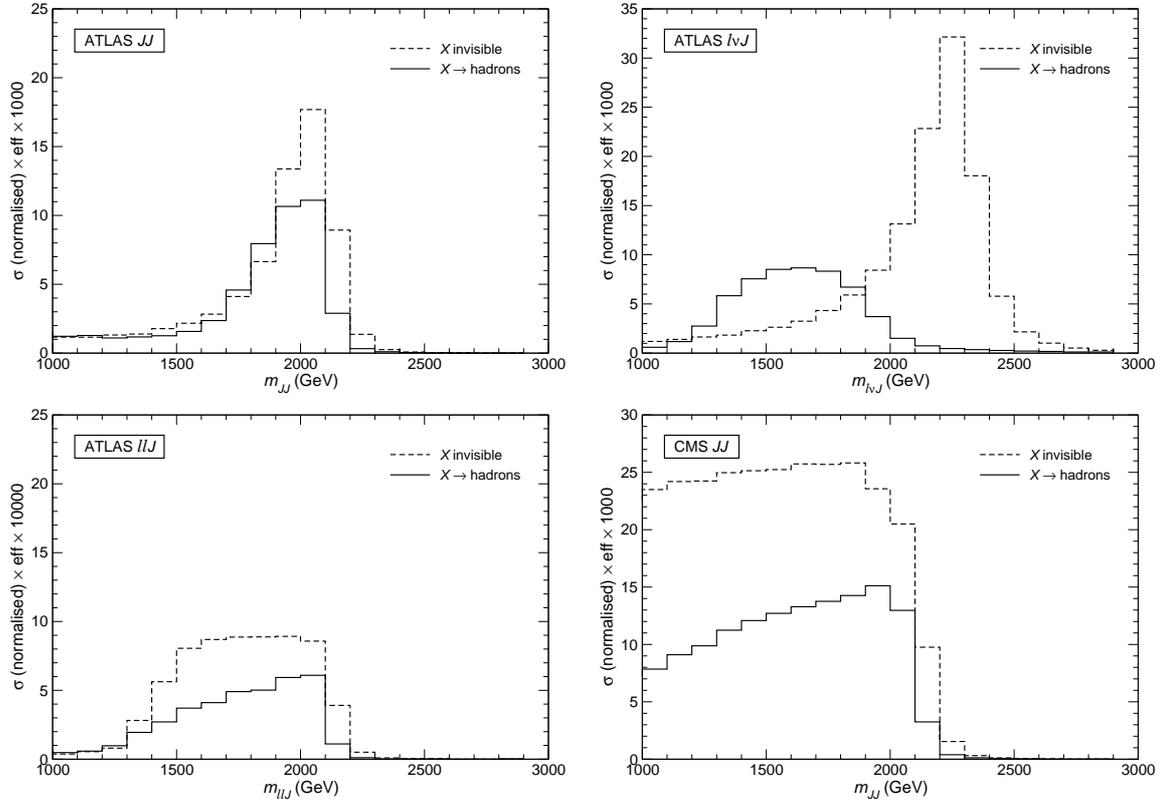

\begin{center}
\begin{tabular}{cc}
\includegraphics[height=5.2cm,clip=]{Figs/mJJ-ATLAS} &
\includegraphics[height=5.2cm,clip=]{Figs/mlnuJ} \\
\includegraphics[height=5.2cm,clip=]{Figs/mllJ} &
\includegraphics[height=5.2cm,clip=]{Figs/mJJ-CMS}
\end{tabular}
\caption{Diboson reconstructed masses in the four analyses considered.}
\label{fig:comp}
\end{center}
\end{figure}

Two benchmarks are used: $X$ invisible and $X$ decaying into two light quarks, with the values for the masses and widths given in section~\ref{sec:2.3} in the former case and $M_R = 2.1$ TeV in the latter. After application of the jet kinematical selection of the ATLAS $JJ$ analysis, the jet mass of the leading and sub-leading jets, respectively labelled as $1,2$, are reasonably well reproduced, see figure~\ref{fig:jetmass}. (The peak at 300 GeV corresponds to the hadronic decay of the boosted particle $Y$.)
We stress that, due to the jet filtering/pruning used by the ATLAS and CMS Collaborations, their jet mass resolutions are considerably better than the one obtained here with the fast detector simulation. Therefore, in this respect our results are conservative and should improve with a more sophisticated analysis. Here, in order to select the $W/Z$ jets we will simply apply a jet mass cut $m_J < 200$ GeV and no $W/Z$ tagging based on jet sub-structure. 
The reconstructed resonance mass is presented in figure~\ref{fig:comp} for the two benchmarks and the four analyses considered. The distributions are normalised to unit cross section before the selection criteria, so that by comparing the four plots one can estimate (up to additional boson tagging efficiency factors) the relative size of the signals in different channels. Let us discuss them in turn.

As expected from the parton-level results, for the ATLAS $JJ$ selection (upper left panel) the peak is clearly visible. Here, it is expected that the use of jet tagging and more stringent mass window cuts would sharpen the peak when $X$ decays hadronically, making it more similar to the peak for invisible $X$, where there is not contamination from extra jets in the events.

In the ATLAS $\ell \nu J$ analysis the diboson resonance mass is reconstructed using the missing energy, assumed to come from the neutrino in the $W$ leptonic decay. If $X$ decays invisibly, it still contributes to the missing energy of the event, therefore the reconstructed diboson mass (upper right panel) sharply peaks at the true resonance mass, assumed $M_R = 2.3$ TeV in this benchmark. Up to different jet tagging efficiency factors for the ATLAS $JJ$ and $\ell \nu J$ analyses (two fat jets in the former and one in the latter, and different mass windows), the height of the two peaks is comparable, and probably such a peak should have been noticed in the $\ell \nu J$ search. Therefore, the scenario of invisible $X$ is disfavoured. Besides, should the ATLAS excess events in the dijet channel have large associated missing energy  (around $200$ GeV in this benchmark), that fact would have been noticed and reported. In case $X$ decays hadronically, the resulting distribution is rather broad and the signal is probably unobservable. This is especially the case if $X$ is the Higgs boson, as both the ATLAS and CMS Collaborations apply a veto on $b$-tagged jets on this channel, in order to suppress the $t \bar t$ background.

In the $\ell \ell J$ final state the reconstructed mass distributions (lower, left panel) are rather flat in both cases. The number of events around 2 TeV is $20$ times smaller than in the dijet channel (notice the different plot scales), thus the signals seem compatible with the small excesses observed in the ATLAS and CMS searches.

Finally, with the CMS dijet selection the mass distributions are again very broad (lower, right panel). The relative size with respect to the ATLAS excess is uncertain, since the jet tagging methods differ and the effiiciencies for this triboson signal are not known. In any case, the flat distributions produced seem compatible with the smaller excess observed by the CMS Collaboration, bearing in mind that the normalisation of the QCD dijet background is done by a fit to data after the event selection and jet tagging are applied.

\section{Conclusions}
\label{sec:4}

A heavy resonance decaying into two massive gauge bosons plus an extra particle might explain the peak-shaped excess in the ATLAS diboson resonance search~\cite{Aad:2015owa} and the absence of such peaks in semi-leptonic channels~\cite{Aad:2015ufa,Aad:2014xka,Aad:2014pha}, nor in the CMS dijet analysis~\cite{Khachatryan:2014hpa}. Simple tests of this hypothesis could be performed by removing the transverse momentum balance requirement in the ATLAS dijet analysis---which would make the excess adopt a broader shape---or, conversely,  by introducing this requirement in the rest of searches, especially in the CMS fat dijet analysis. Dedicated searches, looking for $3J$ resonances, $JJ$ plus additional particles or $JJ$ plus missing energy, would also be welcome.

A question remaining to be answered is the required production cross section. With the efficiency $\sim 0.04$ obtained applying the ATLAS event selection citeria and keeping events around the 2 TeV peak, times the  efficiency $\sim 0.16$ for boson tagging quoted in ref.~\cite{Aad:2015owa}, we estimate that the required signal cross section is 62 fb, somewhat large. To give an example, a new 2 TeV $W'$ boson with coupling $g'=g$ to the right-handed fermions has a total production cross section $\sigma = 53$ fb at leading order, but one also has to include the branching ratios to the desired final state $WZX$.
Model building in this direction is then needed to propose suitable candidates. In this respect, there is considerable freedom because the extra particle $X$ could be neutral or charged, coloured or a colour singlet, and correspondingly there are many possibilities for the heavy resonance $R$, not necessarily produced in quark-antiquark processes. 

Among more exotic candidates, the possibility that $X$ is simply the Higgs boson is quite intriguing. If a $WZH$ resonance $R$ is produced with the above estimated cross section, a 12 fb $WH$ signal will result when the $Z$ boson decays invisibly. In the $W$ leptonic decay mode the invariant mass distribution of the $WH$ pair $m_{\ell \nu J}$ will concentrate around $M_R$, since the invisible $Z$ still contributes to $m_{\ell \nu J}$. (A similar example has already been shown in our analysis of the invisible $X$ scenario, where the distribution in the top right panel of figure~\ref{fig:comp} [$\ell \nu J$ channel] exhibits a peak.) For the hadronic channel there are two possibilities that correspond to the two topologies in figure~\ref{fig:topVY}:
\begin{itemize}
\item For the cascade decay $R \to YZ \to WZH$, the $WH$ invariant mass $m_{JJ}$ will peak at the $Y$ mass $M_Y < M_R$.
\item For $R \to YW \to WZH$, $m_{JJ}$ will be broadly distributed below $M_R$.
\end{itemize}
Therefore, for the topology in figure~\ref{fig:topVY} (b), a peak should manifest in the $WH$ invariant mass distribution in the semi-leptonic channel but not in the fully hadronic one. This is precisely the behaviour suggested by the CMS semi-leptonic~\cite{CMS:2015gla} and fully hadronic~\cite{Khachatryan:2015bma} searches for $WH$ resonances: the former does have a $2.2\,\sigma$ deviation of $\sim 20$ fb at 1.8 TeV whereas the latter, more sensitive, only has an excess at the $1\,\sigma$ level for this mass. Still, one should bear in mind that statistics are not enough to draw any conclusion.

The possibly common origin of the ATLAS $VV$ and CMS $WH$ excesses---where the slight mass differences can be attributed to the energy resolution---certainly deserves a more detailed study of the boosted jet tagging and mass reconstruction of $WZH$ signals. Also, one should bear in mind another $2.8\,\sigma$ excess in final states with two leptons and two jets at an invariant mass of 2 TeV~\cite{Khachatryan:2014dka}, already interpreted as resulting from new $W'$ or $Z'$ vector bosons~\cite{Deppisch:2014qpa,Heikinheimo:2014tba,Aguilar-Saavedra:2014ola}.
Provided the current excesses are confirmed in 13 TeV data, the higher statistics available will allow for exhaustive tests of the various hypotheses of new resonance production.

\section*{Acknowledgements}
I thank S. Rappoccio and M. Pierini for pointing out some inaccuracies in the first version of the manuscript. This work has been supported by MINECO project FPA2013-47836-C3-2-P and by Junta de Andaluc\'{\i}a project FQM 101.

\appendix
\section{Diboson helicities and efficiency in leptonic decays}

For a diboson resonance decay $R \to V_1 V_2$, the possible helicities $(\lambda_1,\lambda_2)$ of the decay products are determined by angular momentum conservation. In the direction of the relative motion of $V_1$ and $V_2$ in the CM frame the orbital angular momentum vanishes, therefore the sum of the spin components in this direction cannot exceed the spin of the resonance. For a scalar $R$ only the like-helicity combinations $(\lambda_1,\lambda_2)=(1,1)$, $(0,0)$, $(-1,-1)$ are allowed. If the scalar has a SM-like coupling $R \, V_{1\mu} V_2^\mu$ the $(0,0)$ helicity combination dominates at large masses, with nearly 100\% of the total $R \to V_1 V_2$ width. For a vector resonance there are four additional combinations allowed, $(\lambda_1,\lambda_2)=(\pm 1,0)$, $(0,\pm 1)$. Again, for a heavy resonance and a SM-like coupling to $V_1 V_2$ the $(0,0)$ combination dominates, altough differences with respect to the scalar case can be found in some spin observables~\cite{Aguilar-Saavedra:2015yza}. The remaining configurations $(\lambda_1,\lambda_2)=(\pm 1,\mp 1)$ imply a total angular momentum of $\pm 2$ in the direction of the relative motion of the decay products, and are possible only for a spin-2 resonance.

\begin{figure}[htb]
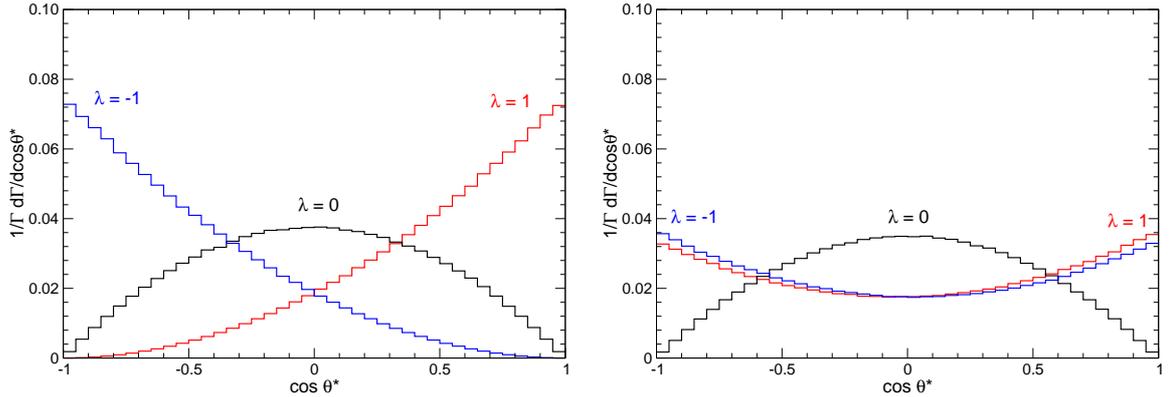

\begin{center}
\begin{tabular}{cc}
\includegraphics[height=5.2cm,clip=]{Figs/costh-Wp} &
\includegraphics[height=5.2cm,clip=]{Figs/costh-Z}
\end{tabular}
\caption{Left: $\ell^+$ distribution in the $W^+$ boson rest frame. Right: $\ell^+$ distribution in the $Z$ boson rest frame.}
\label{fig:ang}
\end{center}
\end{figure}

It is well known~\cite{Kane:1991bg} that in the leptonic decay of a $W$ boson, its helicity determines the charged lepton angular distribution $1/\Gamma \, d\Gamma/d\!\cos\theta^*$, with $\theta^*$ the angle between the charged lepton momentum in the $W$ rest frame and the $W$ boson momentum in the $R$ rest frame. The angular distributions for a $W^+$ boson and its three possible helicity states are shown in figure~\ref{fig:ang} (left). For $W^-$ decays the distributions are the same but with the sign of $\lambda$ interchanged. 
The angle $\theta^*$ subsequently influences the kinematics of the $W$ decay products. The fraction of $W$ energy carried by the charged lepton in the CM frame $E_\ell/E_W$ is
\begin{equation}
\frac{E_\ell}{E_W} = \frac{1}{2} \left[ 1 + \beta \cos \theta^* \right] \,,
\end{equation}
with $\beta$ the $W$ boson velocity, measured in the CM frame. Therefore, $W^+$ bosons with $\lambda=1$ and $W^-$ bosons with $\lambda=-1$ produce high-$p_T$ leptons and small missing energy; on the other hand, $W^+$ bosons with $\lambda=-1$ and $W^-$ bosons with $\lambda=1$ produce much softer leptons and large missing energy. It is then pertinent to ask ourselves about the impact of this difference on the signal efficiency for the $\ell \nu J$ channel. Note that for leptonic decays of the $Z$ boson the differences between the three possible helicities are less pronounced because the coupling to the leptons is almost axial. The distributions for the positively charged lepton are presented in figure~\ref{fig:ang} (right).
In the hadronic $W/Z$ decays the boson helicity affects the jet tagging efficiency, which is slightly larger for $\lambda = 0$ because the jet sub-structure is less visible for $\lambda=\pm 1$~\cite{Khachatryan:2014hpa}.

We estimate the variation in the efficiencies using a fast detector simulation and the event selection criteria for the ATLAS $\ell \nu J$ analysis, collected in section~\ref{sec:3}. We simulate $q \bar q \to R \to W^+ W^-$ samples corresponding to all the helicity combinations, taking $M_R = 2$ TeV, $\Gamma_R = 50$ GeV. (This argument obviously applies to $WZ$ resonances in the $\ell \nu J$ channel too.) All decays of the $W$ bosons are included. We list in table~\ref{tab:effWW} the efficiencies relative to the $(0,0)$ combination that mostly corresponds to a scalar or vector resonance. The largest difference is found for the unlike-helicity combination $(-1,1)$ that produces soft leptons, but this is insufficient to explain the tension between the measurements in the $\ell \nu J$ and $JJ$ channels. We stress that these estimations only take into account the leptonic side of the diboson event; small differences due to the variation in the fat jet tagging are not included. 

\begin{table}[htb]
\begin{center}
\begin{tabular}{c|ccc}
$\lambda_1$ $\backslash$ $\lambda_2$ & 1 & 0 & $-1$ \\ \hline
1 & 0.90 & 0.96 & 0.90 \\
0 & 0.95 & 1 & 0.96 \\
$-1$ & 0.88 & 0.95 & 0.91
\end{tabular}
\caption{Efficiencies for $R \to W^+ W^-$ with the ATLAS $\ell \nu J$ selection, for different $W^+$, $W^-$  boson helicities $(\lambda_1,\lambda_2)$. The values are relative to the combination $(0,0)$. Fat jet tagging efficiencies are not included. \label{tab:effWW}}
\end{center}
\end{table}

A similar analysis can be done for $R \to ZZ$ and the ATLAS $\ell \ell J$ selection criteria. In this case the efficiency variations are rather small, as it is expected from the distributions in figure~\ref{fig:ang}. The largest difference is found for $(\lambda_1,\lambda_2)=(-1,-1)$ with an efficiency a factor of $1.03$ larger than for the $(0,0)$ helicities. The same conclusions apply for $WZ$ resonances in the $\ell \ell J$ channel.


\begin{thebibliography}{99}


\bibitem{Aad:2015owa}
  G.~Aad {\it et al.}  [ATLAS Collaboration],
  arXiv:1506.00962 [hep-ex].

\bibitem{Aad:2015ufa}
  G.~Aad {\it et al.}  [ATLAS Collaboration],
  Eur.\ Phys.\ J.\ C {\bf 75} (2015) 5,  209
  [arXiv:1503.04677 [hep-ex]].

\bibitem{Aad:2014xka}
  G.~Aad {\it et al.}  [ATLAS Collaboration],
  Eur.\ Phys.\ J.\ C {\bf 75} (2015) 2,  69
  [arXiv:1409.6190 [hep-ex]].

\bibitem{Aad:2014pha}
  G.~Aad {\it et al.}  [ATLAS Collaboration],
  Phys.\ Lett.\ B {\bf 737} (2014) 223
  [arXiv:1406.4456 [hep-ex]].

\bibitem{Khachatryan:2014hpa}
  V.~Khachatryan {\it et al.}  [CMS Collaboration],
  JHEP {\bf 1408} (2014) 173
  [arXiv:1405.1994 [hep-ex]].

\bibitem{Khachatryan:2014gha}
  V.~Khachatryan {\it et al.}  [CMS Collaboration],
  JHEP {\bf 1408} (2014) 174
  [arXiv:1405.3447 [hep-ex]].

\bibitem{Khachatryan:2014xja}
  V.~Khachatryan {\it et al.}  [CMS Collaboration],
  Phys.\ Lett.\ B {\bf 740} (2015) 83
  [arXiv:1407.3476 [hep-ex]].
  
\bibitem{deBlas:2012qp}
  J.~de Blas, J.~M.~Lizana and M.~Perez-Victoria,
  JHEP {\bf 1301} (2013) 166
  [arXiv:1211.2229 [hep-ph]].

\bibitem{Aad:2015uka}
  G.~Aad {\it et al.}  [ATLAS Collaboration],
 arXiv:1506.00285 [hep-ex].

\bibitem{Khachatryan:2015bma}
  V.~Khachatryan {\it et al.}  [CMS Collaboration],
  arXiv:1506.01443 [hep-ex].

\bibitem{CMS:2015gla}
  CMS Collaboration,
  Report CMS-PAS-EXO-14-010.

\bibitem{Aad:2015yza}
  G.~Aad {\it et al.}  [ATLAS Collaboration],
  Eur.\ Phys.\ J.\ C {\bf 75} (2015) 6,  263
  [arXiv:1503.08089 [hep-ex]].

\bibitem{Hisano:2015gna}
  J.~Hisano, N.~Nagata and Y.~Omura,
  arXiv:1506.03931 [hep-ph].

\bibitem{Fukano:2015hga}
  H.~S.~Fukano, M.~Kurachi, S.~Matsuzaki, K.~Terashi and K.~Yamawaki,
  arXiv:1506.03751 [hep-ph].

\bibitem{Franzosi:2015zra}
  D.~B.~Franzosi, M.~T.~Frandsen and F.~Sannino,
  arXiv:1506.04392 [hep-ph].

\bibitem{Cheung:2015nha}
  K.~Cheung, W.~Y.~Keung, P.~Y.~Tseng and T.~C.~Yuan,
  arXiv:1506.06064 [hep-ph].

\bibitem{Murayama:1992gi}
  H.~Murayama, I.~Watanabe and K.~Hagiwara,
  Report KEK-91-11.

\bibitem{protos}
J. A. Aguilar-Saavedra.
PROTOS, a PROgram for TOp Simulations. http://jaguilar.web.cern.ch/jaguilar/protos/

\bibitem{pgs4}
PGS 4, see http://www.physics.ucdavis.edu/$\sim$conway/ research/software/pgs/pgs4-general.htm .

\bibitem{Butterworth:2008iy}
  J.~M.~Butterworth, A.~R.~Davison, M.~Rubin and G.~P.~Salam,
  Phys.\ Rev.\ Lett.\  {\bf 100} (2008) 242001
  [arXiv:0802.2470 [hep-ph]].

\bibitem{Ellis:2009su}
  S.~D.~Ellis, C.~K.~Vermilion and J.~R.~Walsh,
  Phys.\ Rev.\ D {\bf 80} (2009) 051501
  [arXiv:0903.5081 [hep-ph]].

\bibitem{Khachatryan:2014dka}
  V.~Khachatryan {\it et al.}  [CMS Collaboration],
  Eur.\ Phys.\ J.\ C {\bf 74} (2014) 11,  3149
  [arXiv:1407.3683 [hep-ex]].
  
\bibitem{Deppisch:2014qpa}
  F.~F.~Deppisch, T.~E.~Gonzalo, S.~Patra, N.~Sahu and U.~Sarkar,
  Phys.\ Rev.\ D {\bf 90} (2014) 5,  053014
  [arXiv:1407.5384 [hep-ph]].

\bibitem{Heikinheimo:2014tba}
  M.~Heikinheimo, M.~Raidal and C.~Spethmann,
  Eur.\ Phys.\ J.\ C {\bf 74} (2014) 10,  3107
  [arXiv:1407.6908 [hep-ph]].

\bibitem{Aguilar-Saavedra:2014ola}
  J.~A.~Aguilar-Saavedra and F.~R.~Joaquim,
  Phys.\ Rev.\ D {\bf 90} (2014) 11,  115010
  [arXiv:1408.2456 [hep-ph]].

\bibitem{Aguilar-Saavedra:2015yza}
  J.~A.~Aguilar-Saavedra and J.~Bernab\'eu,
  arXiv:1508.04592 [hep-ph].

\bibitem{Kane:1991bg}
  G.~L.~Kane, G.~A.~Ladinsky and C.~P.~Yuan,
  Phys.\ Rev.\ D {\bf 45} (1992) 124.

\end{thebibliography}
\end{document}